\documentclass{revtex4}
\usepackage{amssymb}
\usepackage{amsfonts}
\usepackage{amsmath}

\input{tcilatex}

\begin{document}

\title{Ground State H-Atom in Born-Infeld Theory}
\author{S. Habib Mazharimousavi$^{\ast}$}
\author{M. Halilsoy$^{\dag}$}
\affiliation{Department of Physics, Eastern Mediterranean University,}
\affiliation{G. Magusa, north Cyprus, Mersin-10, Turkey}
\affiliation{$^{\ast}$habib.mazhari@emu.edu.tr}
\affiliation{$^{\dagger}$mustafa.halilsoy@emu.edu.tr}

\begin{abstract}
Within the context of Born-Infeld (BI) nonlinear electrodynamics (NED) we
revisit the non-relativistic, spinless H-atom. The pair potential computed
from the Born-Infeld equations is approximated by the Morse type potential
with remarkable fit over the critical region where the convergence of both
the short and long distance expansions slows down dramatically. The Morse
potential is employed to determine both the ground state energy of the
electron and the BI parameter.
\end{abstract}

\maketitle

\section{Introduction}

An intriguing example of nonlinear electrodynamics (NED) was introduced in
1934 by Born and Infeld (BI) in order to eliminate divergences in the
Coulomb problem \cite{1}. With the advent of quantum electrodynamics (QED),
however, divergences were resolved by the well-established scheme of
renormalization. Being popular enough, QED suppressed the efforts of BI to
the extend of being forgotten until very recent decade. We observe now that
BI theory gained recognition anew within string theory; a theory that aims
to unify all forces of nature, including quantum gravity, under a common
title. Once BI paved the way toward NED, various modifications emerged as
alternative theories to the well-known linear electrodynamics of Maxwell.
The common feature in all these NED theories is that in the linear limit it
recovers the Maxwell's electrodynamics, as it should. The NED Lagrangian is
commonly constructed from the invariants $F_{\mu \nu }F^{\mu \nu }$ and $%
F_{\mu \nu }{}^{\star }F^{\mu \nu }$($\star $ means dual) as nonpolynomial
functions. The field tensor $F_{\mu \nu }$ is defined as in Maxwell's linear
electrodynamics, namely, $F_{\mu \nu }=\partial _{\mu }A_{\nu }-\partial
_{\nu }A_{\mu },$ in terms of the vector potential $A_{\mu }.$ Since these
invariants play role in the vacuum polarization, BI theory, or any version
of NED can polarize the vacuum. We note that beside string theory, NED and
naturally BI, find applications in different areas, ranging from black holes
and cosmology to the model of elementary particles\cite{2}. In this paper we
revisit the problem to compute Born-Infeld effects on atomic spectra and
employ a Morse type potential simulation to determine the ground level
energy of the H-atom. We restrict ourselves entirely to the pure electrical
potential (i.e. without magnetic fields) applied to the ground state of the
non-relativistic Schr\"{o}dinger equation and for pedagogical reasons we
shall try to avoid the intricacy of NED as much as we can. Let us note that
throughout our calculations, following Ref. \cite{5} we shall employ
dimensionless variables. We choose $c$ (the speed of light), $e$ (charge of
the electron), $m_{e}$ (mass of the electron) and $\hslash $ all equal to
unity. Accordingly, to recover the dimensionful quantities each variable
should be multiplied with the appropriate factor. To convert the
dimensionless unit of length, for instance, it should be multiplied by the
Compton wavelength $\lambda _{C}=\frac{\hslash }{m_{e}c}.$ Similarly, the
unit magnitude of electric field is converted by the factor $\frac{e}{%
\lambda _{C}^{2}}.$ Conversion of energy is carried out by the factor $%
m_{e}c^{2}$ and so on. The electrostatic BI Lagrangian can be expressed by 
\begin{equation}
L\left( X\right) =\frac{4}{\beta ^{2}}\left[ 1-\sqrt{1+\frac{\beta ^{2}}{2}X}%
\right]
\end{equation}%
where $X=F^{2}=F_{\mu \nu }F^{\mu \nu }$ and $\beta $ is the dimensionless
BI parameter. In the limit $\beta \rightarrow 0$ by applying the L'H\^{o}%
pital's rule\ we recover the Maxwell Lagrangian $L=-X$. The electrostatic
potential is given by%
\begin{equation}
A_{\mu }=\left( V\left( r\right) ,0,0,0\right) =\delta _{\mu }^{0}V\left(
r\right)
\end{equation}%
for an $r$ dependent function $V\left( r\right) .$ The sourceless BI
equation reduces to the single equation 
\begin{equation}
\partial _{\mu }\left( \sqrt{\left\vert g\right\vert }\frac{\partial L}{%
\partial X}F^{\mu \nu }\right) =0
\end{equation}%
in which $\sqrt{\left\vert g\right\vert }$ refers to the square root of
determinant for the spherically symmetric flat metric%
\begin{equation}
ds^{2}=-dt^{2}+dr^{2}+r^{2}\left( d\theta ^{2}+\sin ^{2}\theta \ d\varphi
^{2}\right) .
\end{equation}%
Note that the difference of (3) from the standard linear Maxwell equation is
the appearance of $\frac{\partial L}{\partial X}$ term inside the
parenthesis. As we stated above since the magnetic field is chosen to vanish
our electrodynamics equations consist of (3) alone. Born's solution follows
from (3) which is expressed for convenience in the following form \cite%
{3,4,5} 
\begin{equation}
V\left( \left\vert \mathbf{s}\right\vert \right) =\frac{-\alpha }{\beta }%
\int_{\left\vert \mathbf{s}-\mathbf{s}_{e}\right\vert }^{\infty }\frac{dr}{%
\sqrt{1+r^{4}}}.
\end{equation}%
This expression represents the electrostatic potential at $\mathbf{s}$ due
to a positively charged particle located at $\mathbf{s}_{e}$, and the
constant $\alpha \left( =\frac{e^{2}}{\hslash c}\right) $ stands for the
fine structure constant. Obviously the latter is introduced to prepare the
ground for quantum mechanical treatment. compared with the Coulomb potential 
$V\left( \left\vert \mathbf{s}\right\vert \right) \sim \frac{1}{\left\vert 
\mathbf{s}\right\vert }$, the integral expression of the potential (3) is
complicated enough for an analytic treatment. In the next section we shall
describe the strategy of handling the even more complicated two-body
potential in Born-Infeld theory.

\section{Morse potential versus Born potential in the Hydrogen-atom}

In \cite{5}, Carley and Kiessling addressed (see also \cite{3,4}) the old
problem of computing Born-Infeld effects on the Schr\"{o}dinger spectrum of
the hydrogen atom. They use as Schr\"{o}dinger potential $V\left( r,\beta
\right) $, given by 
\begin{equation}
V\left( r,\beta \right) =-\frac{\alpha }{\beta }\left[ W\left( r/\beta
\right) +\frac{1}{4}B\left( \frac{1}{4},\frac{1}{4}\right) \right] ,
\end{equation}%
in which 
\begin{align}
W\left( r/\beta \right) & =\int_{2\sqrt{2}\beta /r}^{\infty }\frac{f^{\prime
}\left( y\right) }{\sqrt{1+x^{4}}}dx, \\
f\left( y\right) & =\sqrt{\frac{1}{4}+y^{2}-y\sqrt{1+y^{2}}},  \notag \\
\text{(}f^{\prime }\left( y\right) & =\frac{df}{dy}\text{), }r=\left\vert 
\mathbf{s}_{proton}-\mathbf{s}_{electron}\right\vert ,  \notag \\
B\left( .,.\right) & =\text{Euler's Beta function}  \notag
\end{align}%
and $xy=\frac{\beta }{r}.$ Following this expression they showed that in two
different regions, namely for $r>2\sqrt{2}\beta $ and $r<2\sqrt{2}\beta ,$
one finds convergent expansions for $W\left( r/\beta \right) $ as%
\begin{equation}
W\left( r/\beta \right) =\left\{ 
\begin{array}{ccc}
\sum_{k=0}^{\infty }a_{k}\left( r/\beta \right) ^{4k+1} & , & r<2\sqrt{2}%
\beta \\ 
&  &  \\ 
\sum_{k=0}^{\infty }b_{k}\left( \beta /r\right) ^{k} & , & r>2\sqrt{2}\beta%
\end{array}%
\right.
\end{equation}%
in which the first few constants were given explicitly\cite{4}. As a result
of their rigorous analysis it turned out that $r=2\sqrt{2}\beta $ is "the
breakdown point" for both expansions. That is, for small $\frac{r}{\beta 
\text{ }}$ the expansion fails to converge for $r/\beta >2\sqrt{2}$, and for
large $r/\beta $ it fails similarly for $r/\beta <2\sqrt{2}$. \ Getting
closer to $r/\beta $ $=2\sqrt{2}$ requires more terms in both expansions,
which make them unpractical in this region. For practical purposes one of
course can always evaluate the integral (7) by straightforward numerical
quadratures. Yet, whenever two expansions from opposite ends of an interval
fail to overlap in their respective domains of convergence, it is desirable
to have some interpolating formula which bridges the two domains of
convergence. In this note we report on such a simple interpolation,
involving only 4 parameters and the well-known Morse potential \cite{6},
given by (note that the subscript 'a' refers to approximate expression) 
\begin{equation}
W_{a}\left( r/\beta \right) =\left[ G\left( 1-e^{-\kappa \left( r/\beta
-b\right) }\right) ^{2}+V_{\circ }\right]
\end{equation}%
where the dimensionless constants are chosen as%
\begin{align}
G& =-1.8300, \\
V_{\circ }& =0.09805  \notag \\
\kappa & =0.58520  \notag \\
b& =-0.45720.  \notag
\end{align}%
After standard separation of variables, the radial part of the Schr\"{o}%
dinger equation reads, ($s-$state) 
\begin{equation}
-\frac{1}{2r^{2}}\frac{d}{dr}\left( r^{2}\frac{d}{dr}\right) R\left(
r\right) +V_{a}\left( r,\beta \right) R\left( r\right) =\varepsilon R\left(
r\right)
\end{equation}%
where the approximate potential $V_{a}\left( r,\beta \right) $ is obtained
by plugging (9) into (6). We impose\ now the following change of variables 
\begin{equation}
r=\beta \rho \text{ and }u\left( \rho \right) =\rho R\left( \rho \right)
\end{equation}%
to get%
\begin{equation}
-\frac{1}{2}\frac{d^{2}}{d\rho ^{2}}u\left( \rho \right) -\alpha \beta \left[
G\left( 1-e^{-\kappa \left( \rho -b\right) }\right) ^{2}+V_{\circ }+\frac{1}{%
4}B\left( \frac{1}{4},\frac{1}{4}\right) \right] u\left( \rho \right) =\frac{%
\varepsilon }{\alpha ^{2}}\left( \alpha \beta \right) ^{2}u\left( \rho
\right) .
\end{equation}%
The latter equation, by introducing $x=\kappa \left( \rho -b\right) $ yields%
\begin{equation}
-\frac{1}{2}\frac{d^{2}}{dx^{2}}u\left( x\right) -\left\vert A\right\vert
\left( 2e^{-x}-e^{-2x}\right) u\left( x\right) =Eu\left( x\right)
\end{equation}%
where 
\begin{align}
\left\vert A\right\vert & =\frac{\alpha \beta }{\kappa ^{2}}\left\vert
G\right\vert >0,\text{ \ \ } \\
E& =\frac{1}{\kappa ^{2}}\left[ \frac{\varepsilon }{\alpha ^{2}}\left(
\alpha \beta \right) ^{2}+\alpha \beta \left( V_{\circ }+\frac{1}{4}B\left( 
\frac{1}{4},\frac{1}{4}\right) -\left\vert G\right\vert \right) \right] <0. 
\notag
\end{align}%
We notice that since $0<\rho <\infty $ then $\kappa b<x<\infty $ in which
infinity is in contrast with the dimension of the atom. Also we note that
this is (an energy-dependent-potential) Schr\"{o}dinger equation and
therefore in the solution we identify $V_{0}$ (consequently $\alpha \beta $)
and $E$ simultaneously. One can show that the proper solution which
satisfies the boundary conditions%
\begin{equation}
\underset{x\rightarrow \infty }{\lim }u\left( x\right) =0,\text{ \ \ }%
\underset{x\rightarrow \kappa \left\vert b\right\vert }{\lim }u\left(
x\right) =0
\end{equation}%
is given by%
\begin{align}
u\left( x\right) & =C\text{\emph{WhittakerM}}\left( a,\nu ,2ae^{-x}\right) ,
\\
a& =\sqrt{2\left\vert A\right\vert },\text{ \ \ }\nu =\sqrt{2\left\vert
E\right\vert },  \notag
\end{align}%
in which $0<\nu \in 
%TCIMACRO{\U{211d} }%
%BeginExpansion
\mathbb{R}
%EndExpansion
$ and \emph{WhittakerM}$\left( a,\nu ,2ae^{-\kappa \left\vert b\right\vert
}\right) =0$ \cite{7}. In fact here $\nu $ is not a quantum number but it is
a new parameter to adjust the results, and $2ae^{-\kappa \left\vert
b\right\vert }=X_{\nu }$ is the first root of \emph{WhittakerM}$\left( a,\nu
,2ae^{-x}\right) =0.$ It is remarkable to observe that once we choose $\nu $%
, we identify both potential and energy of the system at the same time.

Hence one can show that%
\begin{align}
\frac{\varepsilon }{\alpha ^{2}}& =\frac{1}{\left( \alpha \beta \right) ^{2}}%
\left[ \frac{\kappa ^{2}\nu ^{2}}{2}-\alpha \beta \left( V_{\circ }+\frac{1}{%
4}B\left( \frac{1}{4},\frac{1}{4}\right) -\left\vert G\right\vert \right) %
\right] , \\
\alpha \beta & =\frac{\kappa ^{2}a^{2}}{2\left\vert G\right\vert }.  \notag
\end{align}%
This closed expression helps us to adjust $\nu $ in order to set the
ground-state energy of the $H-$atom in accordance with the empirical data
and consequently to find the corresponding value for Born's parameter $\beta
.$ The following table gives $\alpha \beta $ and three first s-states of a
sample $H-$atom by using the above considerations.\ 
\begin{equation}
\ 
\begin{tabular}{|l|l|l|l|l|}
\hline
& $\alpha \beta $ & $-\frac{\varepsilon _{100}}{\alpha ^{2}}$ & $-\frac{%
\varepsilon _{200}}{\alpha ^{2}}$ & $-\frac{\varepsilon _{300}}{\alpha ^{2}}$
\\ \hline
$\text{Empirical }$ &  & $0.49973$ & $0.12493$ & $0.05553$ \\ \hline
$\text{Our results(Morse-type-potential)}$ & $1.82337$ & $0.49973$ & $%
0.20101 $ & $0.08006$ \\ \hline
\end{tabular}
\tag{Table 1}
\end{equation}%
We plot the exact potential (7) and its Morse-like approximation $%
V_{a}\left( r,\beta \right) =-\frac{\alpha }{\beta }\left[ W_{a}\left(
r/\beta \right) +\frac{1}{4}B\left( \frac{1}{4},\frac{1}{4}\right) \right] ,$
in Fig. 1 with a remarkable fit over the range $\frac{r}{\beta }=0$ to $10$,
bridging the two expansion regions mentioned above. However, note that for
larger $\frac{r}{\beta }$ our interpolation formula does not apply, as the
Morse type potential decays exponentially fast instead of inversely.

\section{Conclusion and discussion}

In conclusion, we recall that our attempt was to show that one can always
find a simulation for the functions whose closed forms are not known. This
provides analytical approximations for the final results. Having such
analytical solutions for physical systems always have some significant
features consisting the application of the results for the similar systems.
In particular we simulated the Born-Infeld-Coulomb (BIC) potential into a
Morse-type potential to find an estimation for the Born's parameter as well
as the accurate ground state energy level. We have shown that in this
approach, Born's parameter has a value between the original value proposed
by Born and the value found by Carley and Kiessling. We must emphasize also
that our Morse simulation is independent of the value of the Born parameter.
It is interesting that a non-linear electrodynamics developed in 1930s by
Born and Infeld can be simulated remarkably by a potential developed in
1920s by Morse which proved to be useful in $2-$atom molecules. With the
difference that the subject now is the H-atom consisted of a fixed proton
and a moving electron. From Table 1 it is seen that for $n=2$ and $n=3$ the
Morse-fitting is inefficient. By invoking a different Born parameter for
each excited state may overcome this discrepancy, however, our concern in
this study is confined to the ground state. Further, since our approach is
non-relativistic with zero angular momentum, consistency with the spectra
obtained in Ref. \cite{8} should not be expected. Our ground state, however,
is consistent with the Schr\"{o}dinger spectrum (with zero angular momentum)
of Ref. \cite{8}, as it should. Morse-simulation of the Dirac equation in
Born-Infeld electrodynamics may be considered as a separate future project.\
Finally we wish to remark that our interest in the Born-Infeld
electrodynamics was aroused while searching for pure electrically charged,
regular black holes in the Einstein-Born-Infeld theory \cite{9}.

\textbf{Acknowledgment:} \emph{We are grateful to Professor M.K.-H.
Kiessling for an illuminating correspondence.}

\textbf{Figure Caption:}

Fig. 1: $\frac{\beta }{\alpha }V\left( r,\beta \right) $ (discrete curve)
and the simulated Morse-like $\frac{\beta }{\alpha }V_{s}\left( r,\beta
\right) $ (continuous curve) versus $\frac{r}{\beta }$. Also in this figure
we give 2-terms (DOT), $3-$terms ( BOX) and $4-$terms (CIRCLE) approximation
for $\frac{\beta }{\alpha }V\left( r,\beta \right) $ given in Ref. \cite{5}
for both regions $r<2\sqrt{2}$ and $r>2\sqrt{2}$, which reveal that once $r$
approaches to $2\sqrt{2}$ from both sides we need more terms to get a good
approximation.

\textbf{Table Caption:}

Comparison of first three s-state energy levels with our
Morse-type-simulated results. It is seen that we obtain the Born parameter
as $\beta =\frac{1.82337}{\alpha }$.

\bigskip

\end{document}